\documentclass[
reprint,
longbibliography,
superscriptaddress,
 amsmath,amssymb,
pra,
noeprint
]{revtex4-2}

\usepackage{graphicx}
\usepackage{dcolumn}
\usepackage{bm}
\usepackage{hyperref}
\hypersetup{
    colorlinks=true,
    allcolors=blue
}
\usepackage[mathlines]{lineno}
\usepackage{siunitx}
\usepackage{csquotes}
\usepackage[english]{babel}
\usepackage{color}
\usepackage{xcolor}

\usepackage[capitalize]{cleveref}
\usepackage[caption=false]{subfig}
\newcommand{\phantomsubfloat}[1]{
    {
        \captionsetup[subfigure]{labelformat=empty}
        \subfloat[][]{#1}
    }%
}
\newcommand{\wcavity}{\omega_0}
\newcommand{\lwcavity}{\kappa_0}
\newcommand{\aop}{\hat{a}}
\newcommand{\aopd}{\hat{a}^\dagger}
\newcommand{\Oext}{\Omega_\text{ext}}
\newcommand{\TLSnum}{N_\mathrm{tot}}
\newcommand{\coupl}{g}

\newcommand{\rhoS}{\hat\rho_S}
\newcommand{\rhoB}{\hat\rho_B}
\newcommand{\Rgg}{ \rho_{gg}}
\newcommand{\Ree}{ \rho_{ee}}
\newcommand{\Rge}{ \rho_{ge}}

\newcommand{\curlyD}{\mathcal{D}}
\newcommand{\bop}{\hat{b}}
\newcommand{\rhoop}{\hat{\rho}}
\newcommand{\Oeff}{\Omega'}
\newcommand{\BoseEinstein}{f}

\newcommand{\gammaP}{\kappa_{\scriptscriptstyle +}}
\newcommand{\gammaM}{\kappa_{\scriptscriptstyle -}}

\newcommand{\cavitypop}{n}

\newcommand{\kappatot}{\tilde\kappa}
\newcommand{\kappat}{\kappa}
\newcommand{\im}{\text{i}}
\newcommand{\hc}{{\text H.c.}}

\newcommand{\fd}[2]{\frac{\text{d} {#1}}{\text{d} {#2}}}
\newcommand{\mv}[1]{ \langle #1 \rangle }
\newcommand{\eqnref}[1]{Eq.~\eqref{#1}}

\newcommand{\spare}[1]{\left[ {#1} \right]}
\newcommand{\cpare}[1]{\left\{ {#1} \right\}}
\newcommand{\wtls}{\omega_\mathrm{TLS}}

\begin{document}

\title{Non-Markovian Effects of Two-Level Systems in a Niobium Coaxial Resonator with a Single-Photon Lifetime of 10 milliseconds}










\author{Paul Heidler}
\thanks{These authors contributed equally to this work}
\affiliation{Institut f\"ur Quantenoptik und Quanteninformation, \"Osterreichische Akademie der Wissenschaften, 6020 Innsbruck, Austria}
\affiliation{Institut f\"ur Experimentalphysik, Universit\"at Innsbruck, 6020 Innsbruck, Austria}

\author{Christian M. F. Schneider}
\thanks{These authors contributed equally to this work}
\email{c.schneider@uibk.ac.at}
\affiliation{Institut f\"ur Quantenoptik und Quanteninformation, \"Osterreichische Akademie der Wissenschaften, 6020 Innsbruck, Austria}
\affiliation{Institut f\"ur Experimentalphysik, Universit\"at Innsbruck, 6020 Innsbruck, Austria}

\author{Katja Kustura}
\affiliation{Institut f\"ur Quantenoptik und Quanteninformation, \"Osterreichische Akademie der Wissenschaften, 6020 Innsbruck, Austria}
\affiliation{Institut f\"ur Theoretische Physik, Universit\"at Innsbruck, 6020 Innsbruck, Austria}

\author{Carlos Gonzalez-Ballestero}
\affiliation{Institut f\"ur Quantenoptik und Quanteninformation, \"Osterreichische Akademie der Wissenschaften, 6020 Innsbruck, Austria}
\affiliation{Institut f\"ur Theoretische Physik, Universit\"at Innsbruck, 6020 Innsbruck, Austria}

\author{Oriol Romero-Isart}
\affiliation{Institut f\"ur Quantenoptik und Quanteninformation, \"Osterreichische Akademie der Wissenschaften, 6020 Innsbruck, Austria}
\affiliation{Institut f\"ur Theoretische Physik, Universit\"at Innsbruck, 6020 Innsbruck, Austria}

\author{Gerhard Kirchmair}
\email{gerhard.kirchmair@uibk.ac.at}
\affiliation{Institut f\"ur Quantenoptik und Quanteninformation, \"Osterreichische Akademie der Wissenschaften, 6020 Innsbruck, Austria}
\affiliation{Institut f\"ur Experimentalphysik, Universit\"at Innsbruck, 6020 Innsbruck, Austria}


\date{\today}

\begin{abstract}
Understanding and mitigating loss channels due to two-level systems (TLS) is one of the main cornerstones in the quest of realizing long photon lifetimes in superconducting quantum circuits. Typically, the TLS to which a circuit couples are modeled as a large bath without any coherence. Here we demonstrate that the coherence of TLS has to be considered to accurately describe the ring-down dynamics of a coaxial quarter-wave resonator with an internal quality factor of $0.5\times10^9$ at the single-photon level. The transient analysis reveals effective non-Markovian dynamics of the combined TLS and cavity system, which we can accurately fit by introducing a comprehensive TLS model. The fit returns an average coherence time of around $T_2^*\approx0.3\,\si{\mu s}$  for a total  of $N\approx10^{9}$ TLS with power-law distributed coupling strengths. Despite the shortly coherent TLS excitations, we observe long-term effects on the cavity decay due to coherent elastic scattering between the resonator field and the TLS. Moreover, this model provides an accurate prediction of the internal quality factor's temperature dependence. 
\end{abstract}

\maketitle


\section{\label{sec:Introduction}Introduction}
High-quality microwave resonators are a promising platform for quantum simulation and quantum information processing. Logical qubits encoded in superconducting cavities demonstrated advantages over transmon qubits in coherence times and information capacity. Specifically, the design of coaxial quarter-wave resonators is widely used because it allows easy integration of qubits, which are a necessity to generate quantum states in the cavity~\cite{chou_Deterministic_2018,Heeres_implementing_2017,ofek_Extending_2016,gao_Entanglement_2019,wang_Efficient_2020}. To be a quantum resource, these cavities crucially require high quality factors at the single-photon level. This is a difficult regime, as research over the past two decades has shown that resonators become sensitive to two-level systems (TLS) sitting in oxide layers on the surface~\cite{muller_understanding_2019,bilmes_Resolving_2020,wenner_Surface_2011,wisbey_Effect_2010,oconnell_Microwave_2008, romanenko_Understanding_2017, wang_Improving_2009,goetz_Loss_2016, kaiser_Measurement_2010,martinis_Decoherence_2005}. Additionally, quasiparticle excitations can be a limiting factor~\cite{goetz_Loss_2016, tinkham_Introduction_2004}.
Due to design optimizations in the participation ratio, state of the art aluminum coaxial quarter-wave resonators achieve internal quality factors as high as $Q_\mathrm{{int}}=10^{8}$ at the single-photon level~\cite{reagor_Quantum_2016, kudra_High_2020}.



Achieving even higher quality factors may require a re-examination of materials. 
Recently, niobium superconducting radio-frequency cavities measured at $T=20 \, \mathrm{mK}$ achieved internal quality factors $Q_\mathrm{{int}} \geq 10^9$ at a power level of about ten photons~\cite{romanenko_ThreeDimensional_2020}. Experiments varying the thickness of niobium oxide layers on the inner cavity surface further demonstrated that TLS in the dielectric metal surface layer affect the internal quality factor~\cite{romanenko_Understanding_2017,verjauw_Investigation_2020}. To further increase coherence times in superconducting circuits it will become increasingly important to better understand the amount of TLS and their properties, such as coherence time and coupling strength. 

Here, we present a niobium coaxial quarter-wave resonator~\cite{reagor_Quantum_2016} with an internal quality factor $Q_\mathrm{{int}}=0.5 \times 10^9$ at the single-photon level. The resonator is fabricated from high-purity 3N5 niobium and treated by buffered chemical polishing to remove the damaged metal surface together with high-loss oxides~\cite{padamsee_RF_2008,jung_Analysis_2014} (see \cref{app:cavity} for details). We use frequency and time-domain measurements to characterize the loss mechanisms in the cavity. Furthermore, we include a transient analysis of the ring up by an extended model including a drive detuning. One main result of this work is that we utilize a comprehensive TLS model, developed in Ref.~\cite{kustura_Effective_2021}, that explains the ring-down behavior of the cavity and consistently predicts the temperature dependence.

 
\section{\label{sec:experimentalapproach}Experimental approach}
\begin{figure*}
    \centering
    \includegraphics[width=17.2cm]{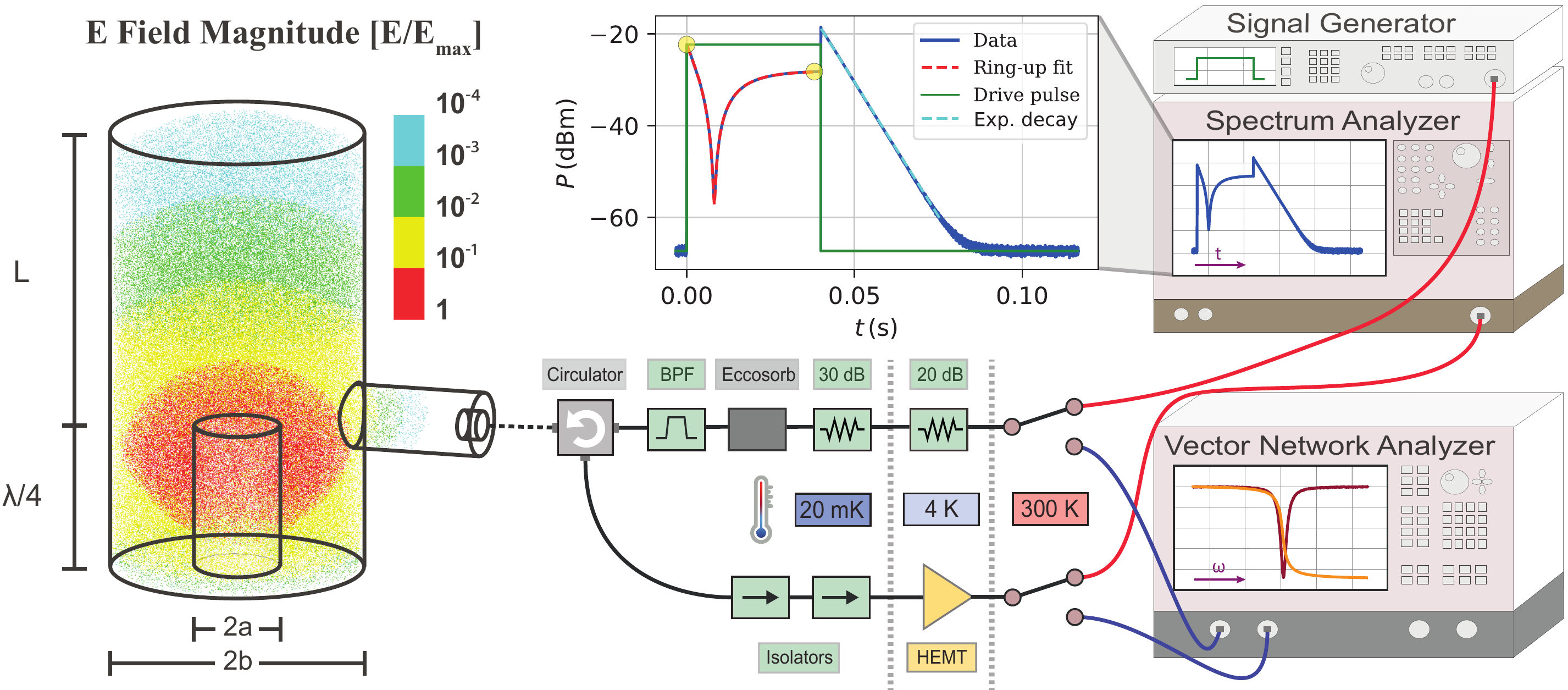}
    \caption{\textbf{High-$Q$ coaxial resonator:} 
   The key element of the particular resonator design is the exponential attenuation of the resonator mode's fields along the circular waveguide section of length $\mathrm{L}$. This is revealed by finite-element simulations where the electric field magnitude is plotted scaled to a peak value of $\mathrm{E_{max}}$ on the tip of the quarter-wave stub. 
    \textbf{Experimental setup:} The cavity is excited either by a cw signal (blue signal path) or a rectangular excitation pulse (red signal path). 
    The input signal enters a cryogenic environment with temperatures down to $T=20\,\mathrm{mK}$ and is routed to the cavity through a coupling pin. The cavity reflected signal is then digitized and analyzed in either the frequency or time domain.
    \textbf{Data acquisition:} The distinctive reflected power response in the time domain of an overcoupled resonator 
    to a resonant rectangular drive pulse is depicted on the screen of the spectrum analyzer. The magnified plot displays additionally the rectangular drive pulse, the cavity ring up fit function, the pure exponential decay and the reflected power levels directly after switching on the cavity drive and after reaching the steady state (yellow marks). The frequency-domain equivalent is shown on the monitor of the vector network analyzer with phase and magnitude of the reflection scattering parameter near the resonance. Both domains allow the extraction of all characteristic parameters.}
    \label{fig:setup}
\end{figure*}

The cavity is mounted in reflection configuration inside a magnetic shield on a dilution refrigerator's base plate, providing a bath temperature of about $20\,\mathrm{mK}$. An illustration of the experimental setup is shown in \cref{fig:setup}. 
A circulator close to the cavity is used to separate the input from the reflected signal, allowing us to use cryogenic amplifiers. The input signal is attenuated by a total of $50~\mathrm{dB}$ and filtered by a $\SI{12}{GHz}$-low-pass filter together with a home-build eccosorb filter. After reflecting off of the cavity, the signal is guided through two 20-dB isolators, a cryogenic amplifier at 4\,K, and another amplifier at room temperature.

For frequency-domain characterizations, we use a standard vector network analyzer (VNA) setup. The VNA operates both as a signal generator and digitizer together with up- and down-mixing. We save the full complex-valued scattering parameters to use a circle-fit routine, allowing the improved extraction of cavity parameters~\cite{probst_Efficient_2015, zoepfl_Characterization_2017}. Typical high and low power measurements are shown in \cref{app:dataprocessing}. The same information can be extracted in the time domain. A signal generator provides a resonant drive pulse, and a spectrum analyzer in zero span mode records the cavity's reflected power response. This is a typical setup for high-$Q$ cavities when the resonance gets too sharp to resolve in the frequency domain~\cite{romanenko_Understanding_2017,padamsee_RF_2008}. In our case, the drive-pulse length is chosen to be $40\,\mathrm{ms}$, which is roughly ten times the cavity lifetime. 


The full transient response of the overcoupled niobium cavity driven by a saturation pulse is shown in the spectrum analyzer illustration in \cref{fig:setup}. Once the pulse arrives at the cavity, the field inside the cavity rings up exponentially. Only a fraction of the traveling signal enters the cavity and is reflected with a phase shift. The other part is reflected directly at the coupling pin, such that the total measured signal is a superposition of two out-of-phase signals. As the energy stored in the cavity rises, the cavity-emitted signal cancels an increasing part of the directly reflected signal. The minimum in reflected power is at the point, where both parts have equal amplitudes. After that, the cavity-emitted signal starts to dominate. Consequently, the response becomes $\pi$ phase shifted (see the time profile of the reflected voltages signal $V_\mathrm{r}$ in \cref{app:dataprocessing}). Since power measurements are phase insensitive, the reflected power rises after the dip until an equilibrium is reached. Note that an ideal critically coupled cavity 
would reflect no signal at all in the steady state since the internal loss rate equals exactly the coupling rate. When the drive is finally turned off, a sudden jump in the reflected power emerges, as one part of the two competing signals vanishes. This is followed by the exponential decay of the cavity population. In our case, the cavity's decay constant increases during the decay, as has been observed before in Ref.~\cite{romanenko_Understanding_2017}.
Understanding the transient behavior allows us to fit time traces for either ring up or ring-down. Ring-up traces would give a high signal-to-noise ratio for the low-photon-number regime if a fast sampling digitizer is used. We do not have such a digitizer available in the experiment and can only acquire traces for a cavity population $\langle n_\mathrm{ph} \rangle > 10^6$. We further assume that the coupling rate is not changing over time or with power. Then we can plot the internal quality factor as a function of average photon number in the cavity.

\section{\label{sec:results}Results} 
\subsection{\label{sec:PowerDependence} Resonator's Power dependence}
\begin{figure*}
    \begin{center}
    \vspace{-2\baselineskip}
    \phantomsubfloat{\label{fig:MainResults_a}}
    \phantomsubfloat{\label{fig:MainResults_b}}
    \phantomsubfloat{\label{fig:MainResults_c}}
    \phantomsubfloat{\label{fig:MainResults_d}}
    \includegraphics[width=\linewidth]{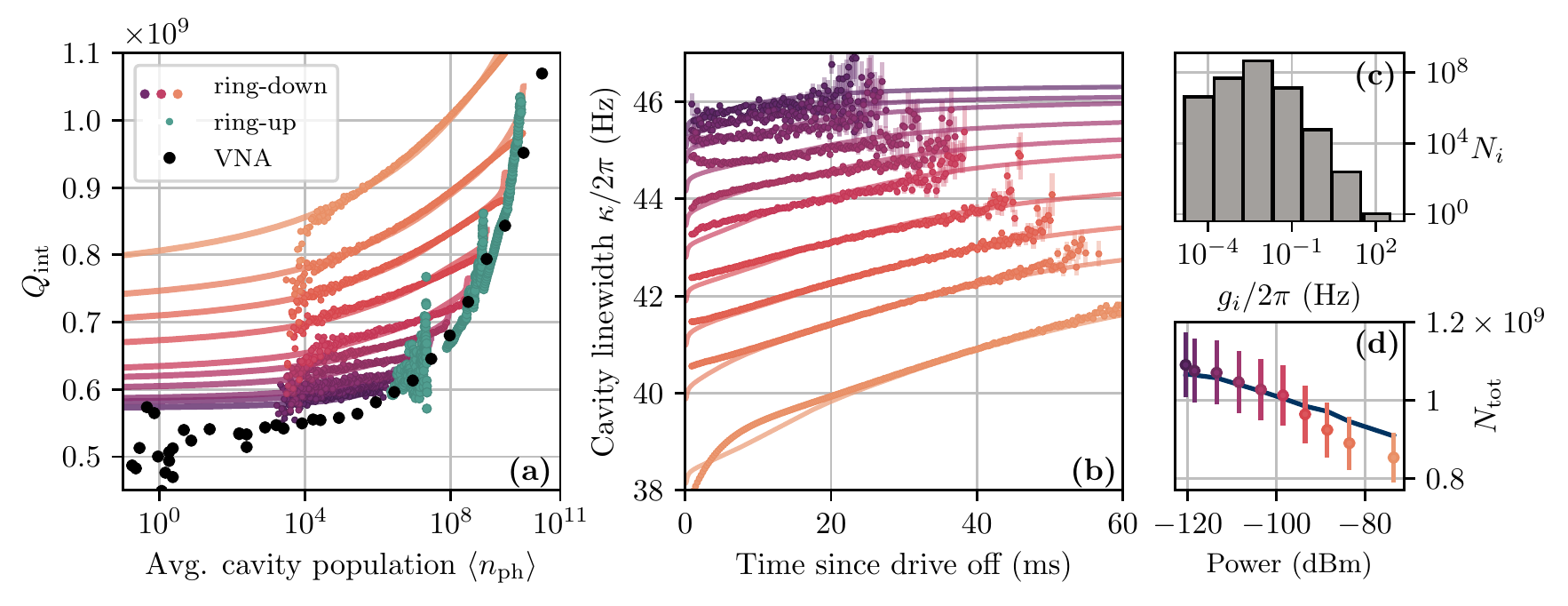}
    \caption{(a) Internal quality factor obtained by steady-state (black), ring-up (mint) and ring-down (purple to orange) measurements as a function of average cavity population. Brighter colors for ring-down traces indicate higher saturation pulse powers, see (d). For readability, we only plot three of the ring-up traces for various pulse powers with the same color. Lower cavity population leads to a lower quality factor. The ring-down measurements clearly differ from the ring-up and VNA measurements. Solid lines show the expected ring-down behavior predicted by the TLS model. (b) Time-dependent cavity linewidth $\kappa$ for the exponential decay of the cavity field after the drive pulse is switched off. Colors are chosen to match (a) and (d). Dots are a rolling average of 50 data points with error bars indicating the standard deviation of the mean ($1\sigma$). The total cavity linewidth is rising as power decays. The prediction of a simultaneous fit of the ten time traces is plotted by solid lines. (c) Sampled TLS distribution for parameters returned by the fit. Only six TLS classes with coupling strengths up to $110\,\mathrm{Hz}$ have a TLS number larger than 1. (d) Total number of TLS returned by the fit as a function of pulse power applied to the cavity port. Colors are chosen to indicate the powers in (a) and (b). The blue line indicates the scaling expected from the linewidth change of the cavity while assuming a constant spectral density of TLS. The error bars indicate 1 standard deviation.
     }
    \label{fig:MainResults}
    \end{center}
\end{figure*}

The evaluated internal quality factor for VNA, ring up and ring-down measurements are shown in \cref{fig:MainResults_a}. One main result of this article is the observed single-photon $Q_\mathrm{int}=5\times10^8$, corresponding to a single-photon lifetime of $\SI{10}{ms}$, which is the highest achieved lifetime for this cavity geometry. 

However, the most prominent feature is that the cavity ring down shows a decreased loss rate as a function of cavity population, compared with the ring-up and steady-state measurements. We are able to predict this behavior through the model of a cavity coupled to $N$ independent TLS, developed in Ref.~\cite{kustura_Effective_2021}. In brief (see \cref{app:TLS} and Ref.~\cite{kustura_Effective_2021} for further details), we first consider a cavity with frequency $\wcavity = 2\pi f_0$ and linewidth $\lwcavity$, subject to coherent resonant driving at a rate $\Oext$, and coupled via the Jaynes-Cummings interaction to a bath of identical TLS with frequencies $\omega_\mathrm{TLS}=\wcavity$ (i.e. TLS resonant to the cavity mode). The time scale at which the TLS dynamics evolve is roughly given by their coherence time $T_2^*$.
Because of the Jaynes-Cummings coupling between TLS and cavity, the TLS experience an effective coherent driving at a rate $g\sqrt{n(t)}$, where $g$ is the coupling rate and $n(t)$ is the cavity occupation at time $t$. This driving adds terms to the TLS dynamics that decay on a time scale $1/\kappa_0$, making them a non-Markovian bath. This non-Markovianity is thus induced through the memory of the cavity itself, and is present even for TLS with very short coherence times.

To obtain the cavity dynamics from the above model, we divide the total time evolution into small steps of duration $\Delta t \gg T_2^*$ and assume the condition $T_2^*\kappa_0\ll1$, so that at each time step the cavity density matrix $\rhoS$ is described by a Markovian master equation,
\begin{multline} \label{cavity-master-equation}
    \fd{\rhoS}{t} = 
    -\im [\Oeff\aop +\hc,\rhoS]
    + \cpare{ \lwcavity\BoseEinstein(\wcavity) + \gammaP }           \curlyD_{\aopd,\aop}[\rhoS]\\
    + \cpare{ \lwcavity\spare{1+\BoseEinstein(\wcavity)} + \gammaM } \curlyD_{\aop,\aopd}[\rhoS],
\end{multline}
with $\hat{a}^\dagger$ ($\hat{a}$) the creation (annihilation) operator of the cavity mode, $\curlyD_{\aop,\bop}[\rhoop] = \aop\rhoop\bop-\{\bop\aop, \rhoop\}/2$ the Lindblad dissipator, and $\BoseEinstein(\omega)=[\exp(\hbar \omega / k_B T)-1]^{-1}$ the Bose-Einstein distribution, with $k_B$ the Boltzmann constant and $\hbar$ the reduced Planck constant. The TLS modify the cavity driving rate to $\Oeff$ and the cavity absorption and decay rates by $\gammaP$ and $\gammaM$, respectively. Crucially all these rates, whose expressions are given in \cref{app:TLS}, are constant within each time step but change from one time step to the next, giving rise to non-Markovian behavior. From the above master equation we can derive an equation of motion for the cavity occupation, $n(t)$ (see \cref{app:TLS}), which we use to fit our experimental data. We emphasize that the above model is an excellent approximation also for TLS out of resonance with the cavity but still within the cavity linewidth, $\vert\omega_\mathrm{TLS}-\wcavity\vert\lesssim \kappa_0$. In this case the relative corrections to the rates $\Oeff$ and $\kappa_\pm$ are of order $(\kappa_0T_2^*)^2$, which are negligible according to the Markov approximation.

Good agreement with our measurements is achieved by adding two refinements to the above model. First, we consider different families of TLS, each coupling to the cavity at a a different rate $g_i$. This is motivated as coupling to TLS is typically assumed to be dipole coupling~\cite{muller_understanding_2019} $\hbar g=\mathbf{d}\cdot\mathbf{E}$, with the electric dipole moment of the TLS $\mathbf{d}$ and the electric field of the cavity mode $\mathbf{E}$. Thus, the coupling rate follows the electric field distribution, which we know from simulations, but may also have a dipole moment distribution, which we do not know. Because of this we use a generic power-law distribution which will be discussed below. 
The second modification is allowing a power-dependent total number of TLS $N_\mathrm{tot}$. As the linewidth of the cavity decreases with increasing power in the cavity, and assuming that TLS frequencies are distributed equally~\cite{muller_understanding_2019}, a narrower linewidth results in fewer TLS effectively being resonant with the cavity. 

In contrast to previous work~\cite{wenner_Surface_2011, wang_Improving_2009}, we get no agreement with measurements when using coupling strengths following the electric field distribution taken from finite-element simulations. In contrast, we use a power-law distribution, similar to Refs.~\cite{martinis_Decoherence_2005, muller_understanding_2019, wenner_Surface_2011, wang_Improving_2009, goetz_Loss_2016}, but motivated by additional different TLS' dipole moments instead of only considering the electric field distribution. Specifically, we use
\begin{equation}\label{distribution}
    \frac{dN(g)}{dg} = \frac{ N_\mathrm{tot} }{\epsilon_s} \frac{1}{1+\left( g/ \epsilon'_s \right)^\beta },
\end{equation}
with the total number of TLS $N_\mathrm{tot}$, a power-law factor $\beta$, and a saturation constant $\epsilon_s$. The distribution is normalized by $\epsilon'_s = \epsilon \beta\sin(\pi/\beta)/\pi$ to give the total number of TLS $N_\mathrm{tot}=\int_0^\infty dgdN(g)/dg$. Note that, in contrast to the typical distribution derived from the standard tunneling model~\cite{martinis_Decoherence_2005, muller_understanding_2019}, this distribution does not diverge for low coupling strengths. 

To compare the prediction from the TLS model to measurements, we convert both the exponential power decay of the measurements and the photon number returned by the model to cavity linewidth over time
\begin{equation}\label{eq:convert_to_kappa}
    \kappa(t) = -\frac{1}{t}\log\left(\frac{P(t)}{P_0}\right)= -\frac{1}{t}\log\left(\frac{n(t)}{n_0}\right).
\end{equation}
Here $P_0$ ($n_0$) is the power (cavity population) at the beginning of the exponential decay at $t=0$ and $P(t)$ ($n(t)$) is the power (cavity population) at time $t$. By using Eq. (3) we do not need to convert detected power to photon number inside the cavity, which would require a characterization of the amplifier chain. Note that this equation is not part of the model describing the cavity decay, it is only used to compare model prediction and measurement data on a logarithmic scale and adjust the error weighting of the least-squares fit. The resulting cavity linewidths for ten acquired time traces with varying pulse powers are plotted in \cref{fig:MainResults_b}. We fit all ten time traces with a total of 1528 data points simultaneously to the full TLS model by allowing 3+10 free parameters, resulting in a reduced chi-squared statistic $\chi^2=0.8$. The prediction of the least-squares minimized model is shown by the solid lines in \cref{fig:MainResults_b} and agrees well with the measurements. The optimized parameters are the average TLS coherence time $T_2^*=(286\pm18)\,\si{ns}$, two parameters characterizing the TLS coupling strength distribution ($\beta=3.26\pm0.01$ and $\epsilon=0.25\pm0.01$), and ten independent $N_{\mathrm{tot},i}$ shown in \cref{fig:MainResults_d}, describing the total number of TLS for each power. Assuming a constant spectral TLS density $dN_\mathrm{tot}/d\omega$, we expect the number of TLS to decrease as the cavity gets narrower $N_\mathrm{tot}=\kappa dN_\mathrm{tot}/d\omega$. This is indicated by the blue solid line in \cref{fig:MainResults_d} and agrees within the error bars to the results from the least-squares fit. Although $T_2^*$ is much smaller than the cavity decay rate, we still get an effect of the TLS at a millisecond time scale because of the effective nonlinear interaction between TLS and the strong cavity field, as explained in the following. 

Importantly, our model returns a Markovian master equation for short time steps, during which the TLS decay much faster than the coupling rate to the cavity mode. The crucial point is that the start condition for each time step is set by the cavity field, inducing a coherent dipole moment on the TLS. This causes coherent scattering of the cavity field on the TLS, essentially creating a mirror. The loss rate of this TLS mirror depends on the cavity population, which leads to an effective memory of the TLS bath on the time scale of the cavity decay. The Markovianity at short time scales allows encoding of these memory effects as effective nonlinearities in a dynamical equation of Lindblad (i.e., Markovian-like) form (see \cref{app:TLS} for a detailed discussion).

For the numerical evaluation we are sampling from \cref{distribution} to get discrete classes of TLS labeled by $i=1,2,3,. . .$. Specifically, we get good agreement with measurements when sampling seven or more classes of TLS within a maximum coupling rate $g_\mathrm{max}=10^3\,\si{s^{-1}}$ and a minimum coupling rate $g_\mathrm{min}=10^{-3}\,\si{s^{-1}}$. For the following, we use seven classes of TLS coupling strengths. \cref{fig:MainResults_c} depicts the sampled distribution for parameters returned by the fit. The probability of strongly coupled TLS decreases proportional to $g^{-3.3}$, resulting in effectively no TLS with a coupling rate higher than 100 Hz. This is in good agreement with results from avoided crossings in qubit experiments that never measured dipole moments above $1\,\si{e\AA}$~\cite{muller_understanding_2019,bilmes_Resolving_2020,lisenfeld_Electric_2019,martinis_Decoherence_2005}, which equals $g\approx110\,\si{Hz}$ for our cavity. This dipole moment was estimated by using finite-element simulations to extract the maximum electric field on the surface of the cavity $E_\mathrm{max}$ at the single-photon power level. 

Furthermore, we can compare the obtained TLS density to qubit-avoided level-splitting experiments~\cite{martinis_Decoherence_2005,muller_understanding_2019}. These experiments are typically limited by the linewidth of the qubit, giving a lower limit on resolvable coupling rates. If we do the same here, and take a lower coupling strength limit $g_\mathrm{min}=0.1g_\mathrm{max}$, we obtain a TLS density of approximately $8300\,\si{GHz^{-1}\,\mu m^{-3}}$ assuming a 3-nm-thick oxide. This is a factor 30 higher than measured densities for aluminum oxide~\cite{lisenfeld_Electric_2019,martinis_Decoherence_2005,muller_understanding_2019}; however, it fits with the fact that the loss tangent of $\mathrm{Nb}_2\mathrm{O}_5$ is typically much higher~\cite{kaiser_Measurement_2010} than that of aluminum oxide. We can estimate the loss tangent as $\tan\delta_\mathrm{TLS} = \sum_i \pi P_i d_i^2/ (3\epsilon_0\epsilon_r)\approx1.5\times10^{-3}$~\cite{martinis_Decoherence_2005,lisenfeld_Electric_2019} with the vacuum and relative permittivity $\epsilon_0$ and $\epsilon_r$, where we estimate the dipole moments of the TLS families as $d_i=g_i/E_\mathrm{max}$, and use the TLS density distribution $P_i=N_i/(\hbar\kappa V_\mathrm{ox})$, with the total cavity linewidth $\kappa$, and the oxide volume $V_\mathrm{ox}$. We chose an oxide thickness of $3\,\si{nm}$, which is the lower bound for optimized niobium~\cite{romanenko_Understanding_2017} and assume that the TLS are distributed equally over the whole cavity surface. The resulting $\delta_\mathrm{TLS}$ agrees well with characterizations of $\mathrm{Nb}_2\mathrm{O}_5$~\cite{kaiser_Measurement_2010}. In contrast, estimating the loss tangent via the electric field participation ratio $0.8\times10^{-7}$ and the internal quality factor for low powers results in $\delta_\mathrm{TLS}=2.2\times10^{-2}$, which is an order of magnitude higher, but close to the values obtained for niobium accelerator cavities~\cite{romanenko_ThreeDimensional_2020}. The factor 10 higher value is consistent with the assumption that the coupling strength distribution not only is following the electric field distribution of the cavity, but also has an additional dependence due to different dipole moments, which is not included in the electric field participation ratio. 

\subsection{\label{sec:Temperature dependence} Resonator's temperature dependence}
\begin{figure}
    \centering
    \vspace{-2\baselineskip}
    \phantomsubfloat{\label{fig:MB_a}}
    \phantomsubfloat{\label{fig:MB_b}}
    \includegraphics[width=\linewidth]{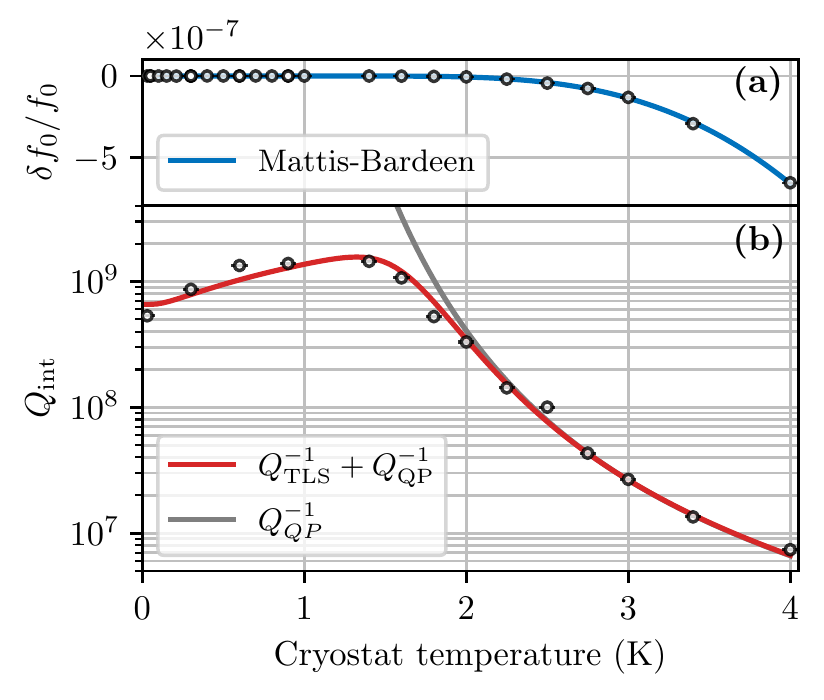}
    \caption{(a) Normalized frequency shift as a function of the cryostat mixing chamber temperature. Increasing temperature increases the kinetic inductance, which translates to a decreased resonance frequency of the cavity. (b) Internal quality factor as a function of temperature. TLS saturation leads to an increase in quality factor for increasing temperature. A plateau arises between 1 and 2\,K due to a temperature-independent dephasing rate of the TLS. For high temperatures we observe an exponential decrease due to quasiparticle excitations. Horizontal error bars indicate the estimated uncertainty for temperature stabilization and vertical error bars the uncertainty returned by the circle-fit routine in both figures. Vertical error bars are smaller than point size. }
    \label{fig:temperature_ramp}
\end{figure}

Finally, we perform a consistency check by measuring the resonance frequency change and internal quality factor of the cavity for increasing mixing chamber temperatures, shown in  \cref{fig:temperature_ramp}. We use low VNA powers to prevent TLS saturation by the cavity field. In contrast to the power-dependent measurements, the coupling distribution should have no impact on the temperature dependence since all TLS saturate the same through the common thermal bath. By fitting Mattis-Bardeen theory (see \cref{app:MB} and Refs.~\cite{gao_Equivalence_2008,mattis_Theory_1958}) to the temperature-dependent frequency shift [see \cref{fig:MB_a}], we extract the kinetic inductance ratio $\alpha=L_\mathrm{kin}/L_\mathrm{tot}=(3.3\pm0.2)\times10^{-5}$, which is the ratio between kinetic inductance $L_\mathrm{kin}$ and total inductance $L_\mathrm{tot}$ of the cavity, and the superconducting gap $2\Delta=(3.06\pm0.03)~\mathrm{meV}$, corresponding to a critical temperature $T_c=(10.1\pm0.1)$\,K. From $\alpha$ we can estimate the skin depth for the niobium cavity $\delta=\mathcal{G}\alpha/\mu_0\omega_0=39.6\,\si{nm}$, which agrees very well with literature values for the London penetration depth of niobium. Here we used finite-element simulations to obtain the geometrical factor $\mathcal{G}=\mu_0\omega_0\int_V |\mathbf{H}|^2 dV/\int_A |\mathbf{J}|^2 dA=74.4\,\Omega$ for our cavity~\cite{reagor_Superconducting_2015,pozar_Microwave_2012}, with the volume integral over the magnetic $\mathbf{H}$ field, the surface integral over the surface current density $\mathbf{J}$, the vacuum permeability $\mu_0$, and the cavity resonance frequency $\omega_0$.

To model the $Q_\mathrm{int}$ dependence on temperature, we sum over two loss channels $Q_\mathrm{int}^{-1} = Q_\mathrm{TLS}^{-1} + Q_\mathrm{QP}^{-1}$. The latter describes resistive losses caused by quasiparticle excitations, for which we use the model~\cite{pozar_Microwave_2012,reagor_Superconducting_2015,tinkham_Introduction_2004}
\begin{equation}
    \label{eq:Q_qp}
    Q_\mathrm{QP} = \frac{\mathcal{G}}{R_\mathrm{sq}} =\mathcal{G}\frac{|\sigma|^2}{\sigma_1\sqrt{(|\sigma|+\sigma_2)\omega_0\mu_0/2}}
\end{equation}
with the temperature-dependent surface resistance $R_\mathrm{sq}$, and the temperature-dependent complex conductivity $\sigma = \sigma_1 - \mathrm{i}\sigma_2$ obtained by Mattis-Bardeen theory (see \cref{app:MB}). By using the value for $\Delta$ obtained by the frequency fit, we only remain with a single free parameter, the normal state conductivity of the cavity surface $\sigma_n$ (hidden in $\sigma$, see \cref{app:MB}).

For TLS losses we use the TLS model from above with the parameters obtained from the time-traces fit. However, we now allow a temperature-independent dephasing time $T_\varphi$ together with a TLS energy relaxation time $T_1$, resulting in the coherence time $T_2^*=\{T_1^{-1}[1/2+f(\omega_0)] + T_\varphi^{-1}\}^{-1}$, with the thermal excitation of the TLS given by the Bose-Einstein distribution $f(\omega)$ from above. In \cref{fig:MB_b}, we show the measurements together with the fit prediction by optimizing the conductivity $\sigma_n=(4.0\pm0.3)\times10^{7}\,\si{S/m}$, $T_1=(723\pm160)\,\si{ns}$, and $T_\varphi=(484\pm111)\,\si{ns}$. We observe good agreement between model and measurements. The resulting $T_2^*=(363\pm65)\,\si{ns}$ is slightly higher than the value returned by the time-trace fit, but the two values agree within their uncertainties. 
Remarkably, we could explain the internal quality factor dependence on temperature completely by TLS coupling and quasiparticle excitations by solely adding a TLS dephasing rate.

 \section{\label{sec:Conclusion}Conclusion}
Summarizing, we present a superconducting niobium quarter-wave coaxial cavity with a single photon lifetime of $\SI{10}{ms}$. The best aluminum coaxial cavities have lifetimes of \SIrange{1}{4}{ms}~\cite{reagor_Quantum_2016, kudra_High_2020} suggesting that switching to niobium cavities can increase the coherence time by a factor of 3 to 10 when taking care of the oxide. Moreover, we extend the ring up formalism to fit the measurements accurately. This is used as a consistency check and, in principle, allows the characterization for low photon numbers when sampling fast enough. 

Our characterizations in both the time and frequency domains indicate that TLS limit our Nb cavity for low photon numbers. However, only the time-domain analysis reveals long-term TLS dynamics that we could explain by coupling the cavity mode to seven classes of TLS with power-law distributed coupling strengths. We show that short-lived TLS can contribute to long-term dynamics due to coherent scattering. This is a very general process, which should be observable in many other systems suffering from TLS loss, e.g. magnons \cite{lachance-quirion_Hybrid_2019}. The model further explains the temperature-dependent losses of the cavity accurately. Understanding and measuring the bath dynamics, coupling rates, and decay times may allow us to better understand TLS and thus improve cavity coherence times in the future. This method can be used to benchmark materials and surface treatments as it allows for a simple characterization of the  TLS on the surface in a more quantitative way. 

The data that support the findings of this work are openly available at Zenodo~\cite{heidler_Dataset_2021}.
 
\begin{acknowledgments}
The authors thank Andreas Strasser for the fabrication of the niobium cavity, and  Gunter Heymann and Gabriel Julian Partl from the Institute of General, Inorganic, and Theoretical Chemistry of the University of Innsbruck for carrying out the niobium cavity buffered chemical polishing. 
P.H. acknowledges support from the Austrian Science Fund FWF within the SFB-BeyondC (Grant No. F7106-N38), C.M.F.S. acknowledges support from the Austrian Science Fund FWF within the DK-ALM (Grant No. W1259-N27) and C. G.-B. acknowledges support from the European Union (PWAQUTEC, H2020-MSCA-IF-2017, Grant No. 796725).
\end{acknowledgments}

\appendix

\section{TLS Model}\label{app:TLS}
We model the cavity dynamics in the presence of TLS impurities using the general TLS bath model developed in Ref.~\cite{kustura_Effective_2021}. The bare cavity (i.e. in the absence of TLS) has a frequency $\wcavity = 2\pi f_0$, linewidth $\lwcavity$, and is subject to coherent resonant driving at a rate $\Oext$. The cavity is coupled via a Jaynes-Cummings interaction to a bath of ${\TLSnum}$ TLS, which we assume to be identical and independent. 
For the sake of illustration, here we assume that the cavity couples to each TLS at the same rate $\coupl$ and that the TLS have a power-independent coherence time $T_2^*=\{T_1^{-1}[1/2 + f(\omega_\mathrm{TLS})] + T_\varphi^{-1}\}^{-1}$, with the TLS energy relaxation time $T_1$, an additional TLS dephasing time $T_\varphi$, and $\BoseEinstein(\omega)=[\exp(\hbar \omega / k_B T)-1]^{-1}$ the Bose-Einstein distribution, with $k_B$ the Boltzmann constant. The extension to TLS-dependent couplings $\coupl_i$ $(i=1,2,3,...)$ is straightforward implemented by summing over the different couplings. 

Since the TLS are identical and independent, their state is a tensor product of single identical TLS  density matrices of the form $\rhoB = \{ \{\Ree,\Rge^*\},\{\Rge,\Rgg\} \}$, whose elements are obtained by solving the optical Bloch equations~\cite{breuer_Theory_2002}. In doing so, we must account for the impact of the cavity occupation, which in this experiment can reach high values, on the TLS dynamics. We model such impact through a coherent TLS driving at a rate $\coupl \sqrt{\cavitypop(t)}$ stemming from the Jaynes-Cummings interaction. Here $n(t) =\langle\hat{a}^\dagger\hat{a}\rangle(t)$ is the cavity occupation with $\aop$ and $\aopd$ the cavity bosonic annihilation and creation operators.
The resulting density matrix $\rhoB(n(t),t)$, and thus the two-time correlators of TLS operators, will in general decay at two different timescales $\tau_{\rm corr,1}$ and $\tau_{\rm corr,2}$. The first is the intrinsic TLS coherence time $\tau_{\rm corr,1} = T_2^*$, and the second is the cavity lifetime $\tau_{\rm corr,2} = 1/\kappa_0$, introduced through the dependence on $\cavitypop(t)$. The cavity dynamics is therefore non-Markovian by definition, as the Markov approximation $\tau_{\rm corr,1},\tau_{\rm corr,2}\ll 1/\kappa_0$ does not hold~\cite{breuer_Theory_2002}.

\begin{figure}
    \centering
    \includegraphics[width=\linewidth]{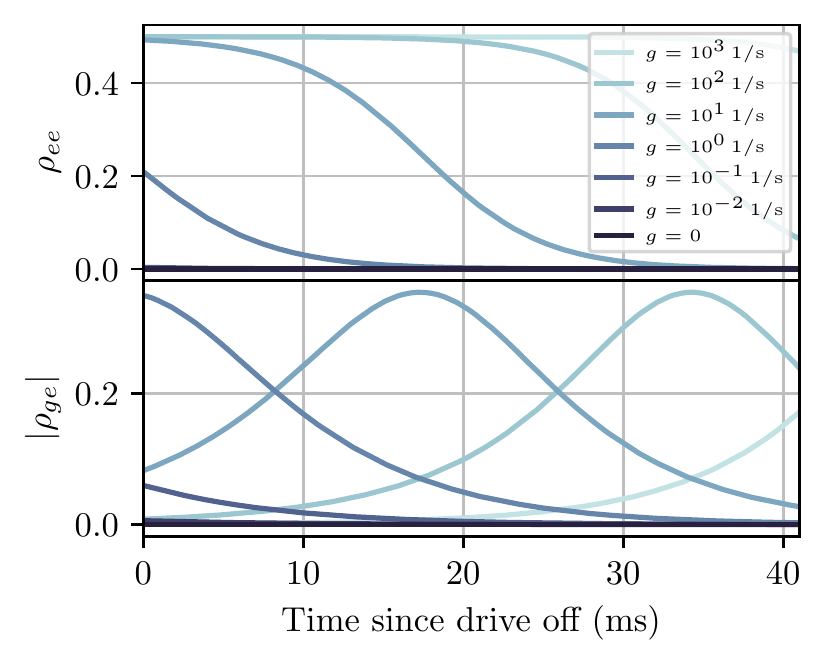}
    \caption{Population and coherences for TLS classes with different coupling strength during the exponential decay for highest used pulse power. Too weakly coupled TLS ($g\le10^{-1}$) do not get populated by the pulse at all. The time domain fit from the main part revealed that there are no TLS with coupling strengths larger than $10^2$. During the slow power decay in the cavity each TLS class has a characteristic time to stop being saturated and build up coherences during the decay.}
    \label{fig:populationandcoherences}
\end{figure}

To compute the non-Markovian cavity dynamics, we further assume $T_2^*\ll1/\kappa_0$. We first focus on a small time interval $[t,t+\Delta t)$ satisfying $T_2^*\ll \Delta t \ll 1/\kappa_0$,
where the cavity occupation can be considered constant, $n(t+\Delta t) \approx n(t)$. Within this interval and in the frame rotating at the frequency $\wcavity$, the cavity density matrix $\rhoS$ obeys the following \emph{Markovian} equation~\cite{kustura_Effective_2021}:
\begin{multline} \label{cavity-master-equation}
    \fd{\rhoS}{t} = 
    -\im [\Oeff\aop +\hc,\rhoS]
    + \cpare{ \lwcavity\BoseEinstein(\wcavity) + \gammaP }           \curlyD_{\aopd,\aop}[\rhoS]\\
    + \cpare{ \lwcavity\spare{1+\BoseEinstein(\wcavity)} + \gammaM } \curlyD_{\aop,\aopd}[\rhoS],
\end{multline}
with $\curlyD_{\aop,\bop}[\rhoop] = \aop\rhoop\bop-\{\bop\aop, \rhoop\}/2$ the Lindblad dissipator, and $\BoseEinstein(\omega)=[\exp(\hbar \omega / k_B T)-1]^{-1}$ the Bose-Einstein distribution, with $k_B$ the Boltzmann constant. The TLS bath induces, first, an additional coherent driving $\Oeff=\Oext+{\TLSnum} \coupl \Rge$ of the cavity stemming from
 elastic scattering of cavity photons. This driving is the most relevant TLS-induced effect~\cite{kustura_Effective_2021} and the main responsible for the observed memory effects.
In addition, the TLS bath induces additional decay and absorption at the rates 
\begin{equation} \label{rates-plus-minus}
    \left(\begin{array}{c} \gammaP  \\ \gammaM \end{array}\right) = 
    \frac{2 \TLSnum \vert \coupl\vert^2 T_2^*}{|\chi|^2}
    \times
    \left(\begin{array}{c}  \Ree - \vert\Rge\vert^2  \\  \Rgg - \vert\Rge\vert^2 \end{array}\right),
\end{equation}
where we define $\chi\equiv 1 + \mathrm{i}(\omega_\mathrm{TLS} - \omega_0)T_2^*$, whose second term describes the correction induced by the finite detuning between TLS and cavity. As the electric field vanishes for frequencies further detuned than the linewidth of the cavity, the coupling rate to these TLS is effectively zero. Therefore, we can give a bound for $|\chi| \le | 1 + \mathrm{i}\kappa_0T_2^*|$, which is close to 1 as $T_2^* \ll 1/\kappa_0$. We therefore assume the TLS within the linewidth to be effectively resonant as corrections are of the order $1/(\kappa_0T_2^*)^2$. The TLS density matrix elements in this time interval correspond to the TLS state at times $t\gg T_2^*$, and read
\begin{equation}\label{TLS-matrix-elements-1}
    \Ree(\cavitypop(t)) = 
        \frac{ |\chi|^2 \BoseEinstein(\wtls) + |g|^2\cavitypop (t)T_1T_2^*/2}
        {|\chi|^2[1 + 2\BoseEinstein(\wtls)] +|g|^2 \cavitypop (t)T_1T_2^* },
\end{equation}
\begin{equation}\label{TLS-matrix-elements-2}
    \Rge(\cavitypop(t)) = 
    \frac{ \im \chi \coupl^*\sqrt{\cavitypop(t)}T_2^*}
    {\vert\chi\vert^2[1+2\BoseEinstein(\wtls)] + |g|^2 \cavitypop (t) T_1T_2^* },
\end{equation}
and $\rho_{gg}=1-\rho_{ee}$.
We see that we can saturate the TLS either by a high thermal population $f(\omega)\rightarrow\infty$ or by a high cavity population $n(t)\rightarrow\infty$. In \cref{fig:populationandcoherences} we plot \eqnref{TLS-matrix-elements-1} and \eqnref{TLS-matrix-elements-2} as a function of time during the cavity exponential decay, for different coupling strengths $\coupl$, and for the highest pulse power of the experiment.

Using the above results we can iteratively solve the cavity dynamics.
First, using \eqnref{cavity-master-equation} we obtain dynamical equations for the expectation values $\mathbf{a}=( \cavitypop,\mv{\aop}, \mv{\aopd})^T$ in an arbitrary time interval of length $[t,t+\Delta t)$, which form a closed  system:
\begin{equation}\label{ada-system}
\dot{\mathbf{a}} =A_S  \mathbf{a}+\mathbf{v}_S,
\end{equation}
with
\begin{equation} \label{ada-system-matrix}
A_S =
\begin{pmatrix}
    -\kappatot  &\im\Oeff       &-\im\Oeff^*    \\
    0           &-\kappatot/2   &0              \\
    0           &0              &-\kappatot/2   \\
\end{pmatrix}, \quad
\mathbf{v}_S =\left(\begin{array}{c} \gammaP  \\ -\im \Oeff^* \\ \im \Oeff \end{array}\right),
\end{equation}
and $\kappatot=\lwcavity +\gammaM-\gammaP$. Note that the coefficients $A_S$ and $\mathbf{v}_S$ are constant within the chosen interval, but their value changes from one interval to another as they depend on $n(t)$.
To compute the cavity dynamics on a time interval $t\in[0,t_f]$, with $t_f$ larger or comparable to $1/\kappa_0$, we divide such interval into $M$ discrete steps $t_1=0, t_2, t_3,...,t_M=t_f$ of length $\Delta t = t_f/(M-1)\ll1/\kappa_0$. 
The value of $\mathbf{a}$ in the time interval $[t_{i-1},t_i)$, namely $\mathbf{a}_{i}$, is obtained recursively using previous values and the analytical solution of \eqnref{ada-system}. Specifically,  $\mathbf{a}_{i} = e^{A_{S,i-1} \Delta t} (\mathbf{a}_{i}-\mathbf{a}_{ss,i})+\mathbf{a}_{ss,i}$, with $\mathbf{a}_{ss,i} = -A_{S,i-1}^{-1}\mathbf{v}_{S,i-1}$. Note that, provided $M$ is large enough to ensure convergence, this method is equivalent to solving, via the discrete Euler method, the \emph{nonlinear} equation
\begin{equation}\label{ada-system2}
\dot{\mathbf{a}} =A_S (\mathbf{a}) \mathbf{a}+\mathbf{v}_S(\mathbf{a}).
\end{equation}
The non-Markovianity or memory effects thus manifests as an effective dynamical nonlinearity for the cavity. Note that this is not always the case, and is a consequence of (i) the non-Markovianity stemming from the back-action of the system itself (i.e. from the slow cavity relaxation), and (ii) the convenient separation of timescales $T_2^*\ll1/\kappa_0$.
For a ring up configuration we set $\Oext\ne0$ and the initial condition $\mathbf{a}_{0} = (0,0,0)^T$ in our iterative algorithm. Conversely for a ring-down configuration, where a driving $\Oext$ is applied at $t<0$ and turned off at $t=0$, we set $\Oext=0$ and take as initial condition $\mathbf{a}_{0}$ the cavity steady state under the driving $\Oext$. Such initial condition is obtained by solving the nonlinear system of algebraic equations $[A_S(n(0))]^{-1}\mathbf{a}_{0}=\mathbf{v}_{S}(n(0))$. For the ring-down in the resonant case, we can give an explicit equation for the change in cavity photon number for sufficiently small time steps ($T_2^* \ll \Delta t \ll 1/\kappa_0$)  
\begin{equation}
\begin{aligned}
n(t_i) &= a+b e^{-\kappatot(t_{i-1}) \Delta t}+c e^{-\kappatot(t_{i-1}) \Delta t / 2}, \\
a &=\frac{\kappa_{+}(t_{i-1})}{\kappatot(t_{i-1})}+\frac{4\left|\Oeff\right|^{2}}{\kappatot^{2}(t_{i-1})}, \\
c &=\frac{4}{\kappatot(t_{i-1})} \operatorname{Re}\left[\mathrm{i} \sqrt{n(t_{i-1})}\Oeff\right]-\frac{8\left|\Oeff\right|^{2}}{\kappatot^{2}(t_{i-1})}, \\
b &=n(t_{i-1}) - a - c, \\
\Oeff &= \TLSnum \coupl \Rge(t_{i-1}).
\end{aligned}
\end{equation}
Finally, after numerically solving the full time dynamics we fit the resulting cavity occupation to the function $\cavitypop(t) = \cavitypop(0) e^{-\kappat(t) t}$, where $\kappat(t)$, namely the time-dependent cavity linewidth, is displayed for a ring-down measurement in \cref{fig:MainResults_b}.

\section{quasiparticle model (Mattis-Bardeen theory)} \label{app:MB}
The current response ($\mathbf{J}$) of a superconductor to an applied time varying electric field ($\mathbf{E}(\omega, t)$) in the presence of quasiparticle excitations can be modeled by a complex conductivity
\begin{equation}
   \mathbf{J}(\omega, t, T) = \left[\sigma_1 - i\sigma_2\right]\mathbf{E}e^{-i\omega t}.
\end{equation}
\textit{Mattis and Bardeen} derived  a thorough relation for the real and imaginary part of the conductivity $\sigma= \sigma_1 - i\sigma_2$ based on BCS theory \cite{mattis_Theory_1958}. In the dirty limit and for $\hbar\omega<2\Delta$ one can simplify their result to \cite{gao_Equivalence_2008,rotzinger_Aluminiumoxide_2016a, tinkham_Introduction_2004}
\begin{subequations}
	\label{eq:cqem:mattisbardeen}
	\begin{align}
		\begin{split}
			\frac{\sigma_1}{\sigma_n} = \frac{4\Delta}{\hbar\omega}e^{-\frac{\Delta}{k_BT}}\sinh\left(\frac{\hbar\omega}{2k_BT}\right)K_0\left(\frac{\hbar\omega}{2k_B T}\right),
		\end{split}\\
	\begin{split}
		\frac{\sigma_2}{\sigma_n} = \frac{\pi\Delta}{\hbar\omega}\tanh\left(\frac{\Delta}{2k_B T} \right),
	\end{split}
	\end{align}
\end{subequations}
with the modified Bessel function of the second kind $K_0(x)$ and a temperature dependent gap $\Delta=\Delta(T)$. We evaluate this using the approximation \cite{gao_Equivalence_2008}
\begin{equation}
	\Delta \approx \Delta_0 \left( 1 - \sqrt{\frac{2\pi k_B T}{\Delta_0}}e^{-\frac{\Delta_0}{k_B T}}\right),
\end{equation}
which is valid for $k_BT\ll \Delta$. With this we can fit the temperature dependent normalized frequency shift
\begin{equation}
	\frac{f_r(T) - f_r^0}{f_r^{0} }  = \alpha\frac{\sigma_2(T) - \sigma_2(0)}{2\sigma_2(T)}.
\end{equation}
Here $\alpha=L_\mathrm{kin}/L_\mathrm{tot}$ is the kinetic inductance ratio. Note that the denominator is actually $2\sigma_2(T)$, and not T=0. We fit the internal quality factor change with \cref{eq:Q_qp} of the main text. 

\section{Data processing}\label{app:dataprocessing}

\subsection{Frequency domain analysis}
\begin{figure}
    \centering
    \includegraphics[width=\linewidth]{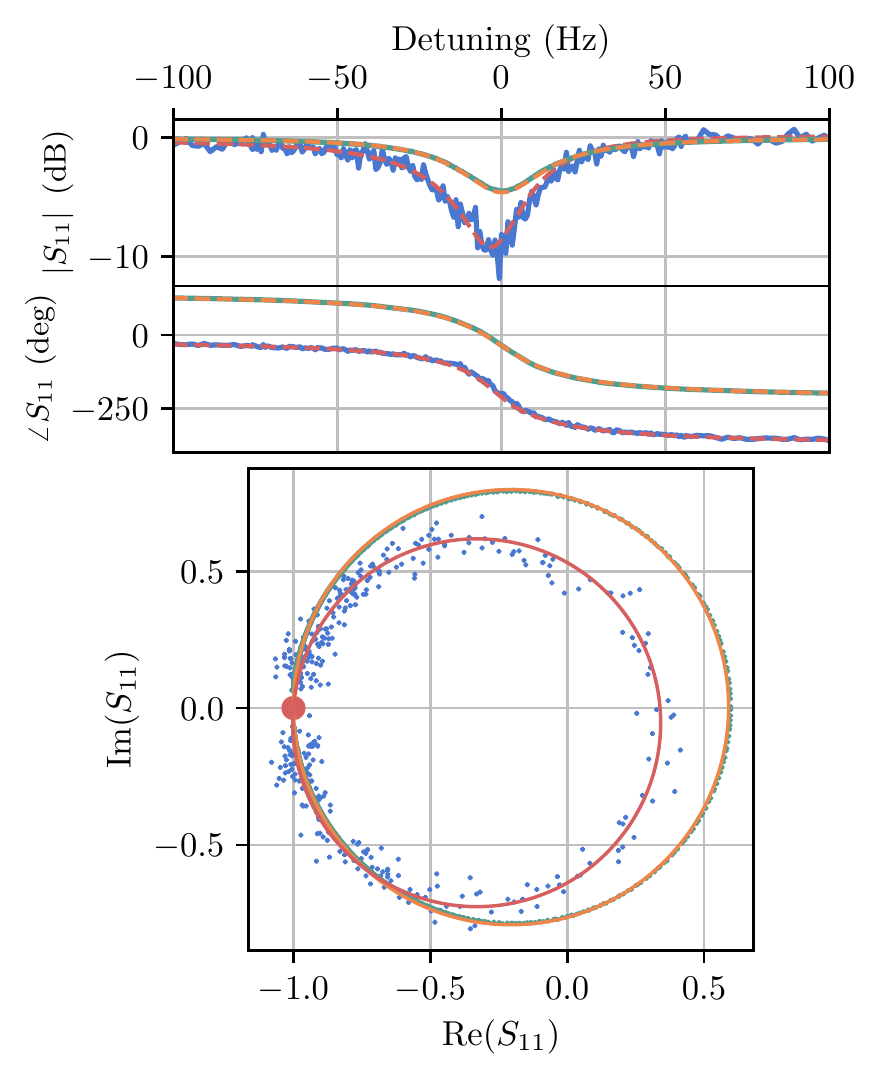}
    \caption{Circle-fit for single-photon power level (data:blue, fit:red) and for $10^{10}$ photons (data:green, fit:orange) VNA measurements. The top two plots show magnitude and phase of the reflected signal as a function of frequency. The bottom plot shows the imaginary and real plane of the complex scattering parameters, together with the fitted circle. An increase in radius indicates increasing internal quality factor, if the coupling stays the same. The internal quality factor rises here from $5.3\times10^{8}$ to $9.4\times10^{8}$ for a 100 dB increase in power.}
    \label{fig:circlefit}
\end{figure}

We analyze complex-valued scattering parameters VNA traces by a circle fit routine to extract resonance frequency, impedance mismatch, internal quality factor $Q_\text{int}$ and coupling quality factor $Q_c$~\cite{probst_Efficient_2015, zoepfl_Characterization_2017}. The magnitude, phase, and circle-fits of the reflection scattering parameters for low and high input drive power levels are shown in \cref{fig:circlefit}. For powers at the single-photon level, we measure a lifetime $\tau_\mathrm{int}=10\,\mathrm{ms}$, the highest achieved lifetime for coaxial quarter-wave resonators to our knowledge. Increasing the power reveals that $Q_\mathrm{int}$ increases to above $1.2 \times 10^9$ for high powers. This power-dependent rise is typical for saturable TLS, which typically sit on the oxide surface of the cavity~\cite{romanenko_Understanding_2017,romanenko_ThreeDimensional_2020, kaiser_Measurement_2010, pappas_Two_2011,oconnell_Microwave_2008,barends_Contribution_2008}. 

\subsection{Time-domain analysis}\label{app:timedomain}
\begin{figure}
    \centering
    \includegraphics[width=\linewidth]{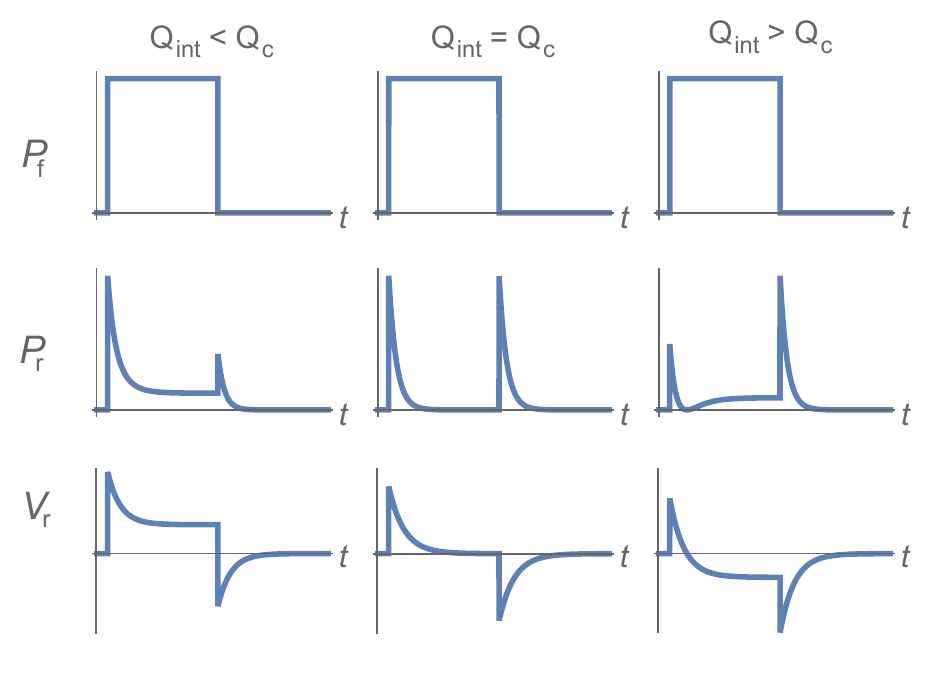}
    \caption{Schematic cavity response to a resonant rectangular pulse $P_\text{f}$ in the under-, critically- and over-coupled regime. $P_\text{r}$  is the reflected power signal and $V_\text{r}$ the reflected wave amplitude. Adapted from \cite[Figure 8.5]{padamsee_RF_2008}.}
    \label{fig:diff coupling regimes}
\end{figure}
\begin{figure}
    \centering
    \includegraphics[width=\linewidth]{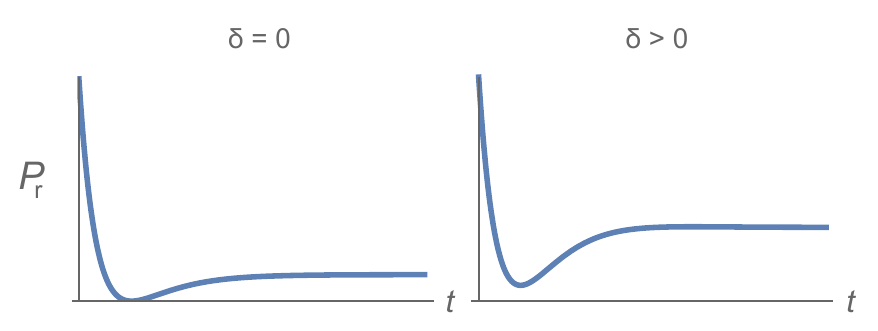}
    \caption{Transient cavity behaviour in the over-coupled regime after switching on a resonant (left) or slightly detuned (right) cavity drive. $P_\mathrm{r}$ describes the cavity's reflected power, $\delta$ the detuning between the cavity drive and the resonance frequency. Except for $\delta$, all other parameters are identical in both cases.}
    \label{fig:drivedetuning}
\end{figure}

We can fit ring up or ring-down data to characterize the cavity. The derivation of the analytic formula for the cavity ring up is done via Laplace- and inverse Laplace-transform. In the case of a resonator in reflection configuration, the transfer function can be written in terms of the complex parameter  $s=i\omega$, with the stored energy in the cavity to be zero at $t=0$, as
\begin{equation}
    \label{transfer f s}
    h(s)=\frac{2Q_\mathrm{l}/Q_\mathrm{c}}{1+\frac{2Q_\mathrm{l}s}{\omega_0}-2iQ_\mathrm{l}}-1,
\end{equation}
with the loaded quality factor $Q_\mathrm{l}$, the coupling quality factor $Q_\mathrm{c}$ and the angular resonance frequency $\omega_0$.
The resonant signal pulse in the time domain starting at $t=0$ is modeled with
\begin{equation}
    \label{input signal det}
    V_{in}(t)=V_\mathrm{f}\,\Theta(t)\big( \cos{(\omega' t) }+i\sin{(\omega' t)} \big)
\end{equation}
$\Theta(t)$ depicts the Heaviside step function, $V_\mathrm{f}$ the voltage amplitude and  $\omega'=2\pi(f_0+\delta)$ the angular resonance frequency including a small drive detuning. The expression for the reflected signal $V_{r}(t)$ in the time domain follows
\begin{equation}
\label{Convolution angewand}
V_{r}(t)= \mathcal{L}^{-1} \lbrace {\mathcal{L}\lbrace V_{in}(t) \rbrace  h(s)}  \rbrace,
\end{equation}
where $\mathcal{L}^{-1}$ depicts the inverse Laplace transform.
The cavity reflected power then reads as

\begin{multline}
    \label{bulky laplace}
    P_{r}(t) = |V_{r}(t)|^2=\frac{P_\mathrm{f}\,}{{\left(Q_\mathrm{c}+Q_\mathrm{int}\right)}^2 f_0^2+{\left(2Q_\mathrm{c} \delta Q_\mathrm{int}\right)}^2} \\
    \Big[{\left(2Q_\mathrm{c} Q_\mathrm{int} \delta\right)}^2+4\operatorname{exp}\left(-2\frac{\left(Q_\mathrm{c}+Q_\mathrm{int}\right) \pi f_0 t}{Q_\mathrm{c} Q_\mathrm{int}}\right) Q_\mathrm{int}^2 f_0^2+\\
    {\left(Q_\mathrm{c}-Q_\mathrm{int}\right)}^2 f_0^2+
    4Q_\mathrm{int} f_0 \operatorname{exp}\left(-\frac{\left(Q_\mathrm{c}+Q_\mathrm{int}\right) \pi f_0 t}{Q_\mathrm{c} Q_\mathrm{int}}\right)\\  \left(\left(Q_\mathrm{c}-Q_\mathrm{int}\right) f_0 \cos\left(2\pi \delta t\right)
    -2Q_\mathrm{c} Q_\mathrm{int} \delta \sin\left(2\pi \delta t\right)\right)\Big],
\end{multline}
with the input signal power $P_\mathrm{f}=|V_{f}|^2$.\par
The reflected voltage and power ring up signals for different coupling regimes are shown in \cref{fig:diff coupling regimes}. The effect of a slightly detuned cavity drive is depicted in \cref{fig:drivedetuning}.\par
\begin{figure}
    \centering
    \includegraphics[width=\linewidth]{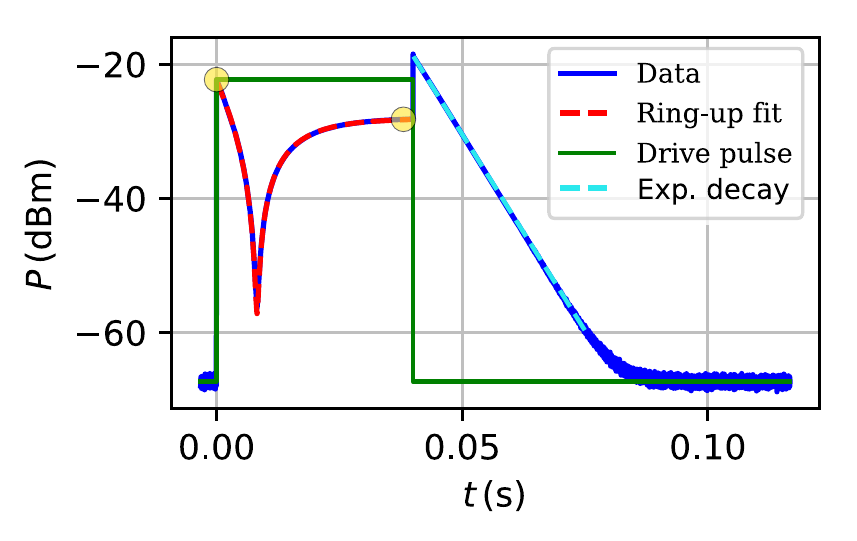}
    \caption{Cavity reflected power trace in the time domain (blue) responding to a resonant rectangular drive pulse (green). Moreover, the cavity ring up fit function (red dashed line) and a pure exponential decay (cyan dashed line) is shown.  The decay constant, combined with the reflected power levels directly after switching on the cavity drive and after reaching steady state (yellow marks), as well as the ring up fit lead to all relevant resonator parameters. Additionally, the ring up fit detects a cavity drive detuning, as the minimum of the transient response gets lifted from the noise floor.}
    \label{fig:Time domain analysis}
\end{figure}

\begin{figure}
    \centering
    \includegraphics[width=\linewidth]{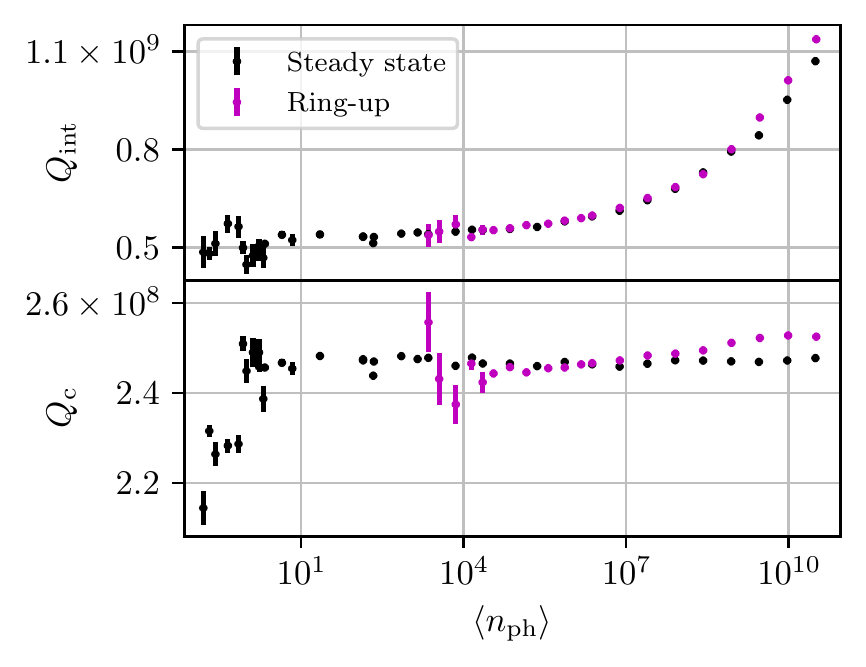}
    \caption{Power dependence of coupling and internal quality factors obtained from steady state and ring up measurements at different cavity drive strengths in units of photons $\langle \mathrm{n_{ph}} \rangle$. We fit a full circle to VNA scattering parameter data in the complex plane to extract all relevant resonator parameters in the frequency domain. In the time domain, the distinctive reflected power ring up trace is fitted with an equivalent, analytically derived model. As we fit one curve to the whole ring up, the quality factor results can be interpreted as averaged values, linked to the maximum of the inner cavity circulating photon number after reaching steady state.}
    \label{fig:Q-factor-comp}
\end{figure}

 
By fitting the full model, we can extract all the relevant cavity parameters, including a drive detuning. The fact that the reflected power minimum in \cref{fig:Time domain analysis} is not reaching the noise floor indicates a small cavity drive detuning. This could be explained due to a frequency shift induced by TLS saturation. The fit results for the detuning are between $0.5-1.1 \, \mathrm{Hz}$ for different input powers. A comparison of the fit parameters $Q_\mathrm{int}$ and $Q_\mathrm{c}$ extracted from either frequency domain VNA measurements or reflected power ring up data is shown in \cref{fig:Q-factor-comp}. One difficulty arises, as the power dependent $Q_\mathrm{int}$ changes during the ring up. This causes slightly  enhanced   $Q_\mathrm{int}$ and $Q_\mathrm{c}$ fit results at high input power levels. The problem of an increasing $Q_\mathrm{int}$ due to TLS saturation after switching on the cavity drive can be addressed by segmented ring up fits.


Another way to evaluate $Q$-factors is to obtain the exponential decay constant after switching off the cavity drive. The reflected power levels of the steady-state and directly after switching on the cavity drive pulse give $Q_\mathrm{int}$ and $Q_\mathrm{c}$. We can infer the total cavity linewidth by \cref{eq:convert_to_kappa} from the main text. From this we can infer the internal quality factor by subtracting the coupling rate $\kappa_c$ and using $Q_\mathrm{int}=\omega/(\kappa(t) - \kappa_c)$.

Both ways of characterizing the cavity's performance in the time domain confirm the high quality factors of the niobium coaxial quarter-wave resonator measured in the frequency domain. In contrast to ring-down measurements, which vanish in the noise floor, acquisitions with a high sampling rate of the ring up would allow to investigate the cavity behavior for low photon numbers at the very beginning of the pulse.

\section{Cavity details}\label{app:cavity}

\subsection{Cavity design and surface treatment}
We use a coaxial quarter-wave cavity fabricated from high-purity 3N5 niobium with a resonance frequency $f_{0}=7.9 \, \mathrm{GHz}$. The radii $a = 2.2 \, \mathrm{mm}$ and $b = 5.8 \, \mathrm{mm}$ are chosen to reach a $3 : 1$ ratio after surface removal minimizing the loss in the resonator. The waveguide section amounts to $L = \SI{36}{mm}$ and the quarter-wave stub is \SI{8}{mm} long. 
To remove the damaged surface layer we use buffered chemical polishing (BCP). The resonator is placed into \SI{555}{ml} of an approximate $1 : 1 : 1$ BCP mixture inside a Teflon container. We keep the BCP mixture during the whole etching process at a temperature of roughly $0-5 \, ^\circ\text{C}$. After 1h, phosphoric acid is added gradually in order to switch to the more moderate BCP composition of $1 : 1 : 2$. A magnetic stirrer is present during the whole process aiming for a steady flow of the BCP mixture. After a total treatment time of 2h, we take out the niobium resonator and rinse it with de-ionized water and acetone. The total surface removal amounts to about $150\,\mu \mathrm{m}$.

\subsection{Aging}
\begin{figure}
    \centering
    \includegraphics{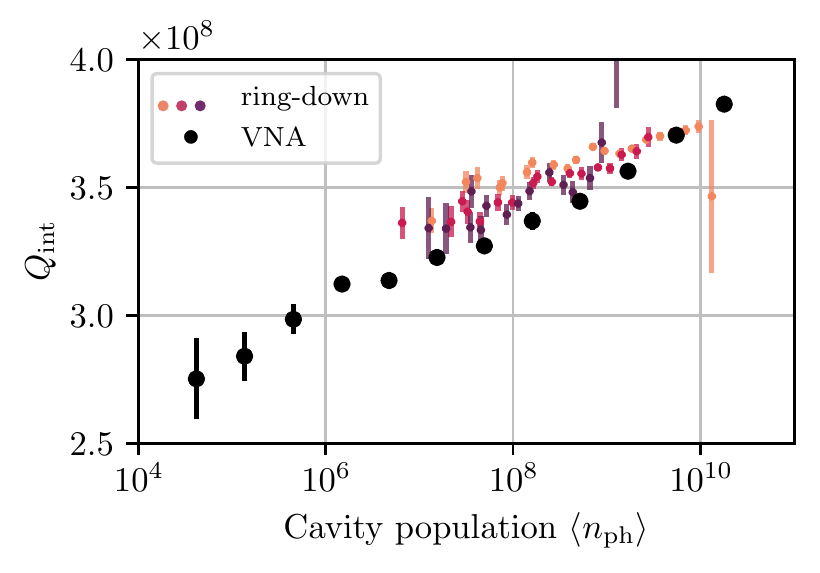}
    \caption{$Q_\text{int}$ power dependence of the aged niobium coaxial quarter-wave resonator as a function of power in cavity in units of photons $\langle \mathrm{n_{ph}} \rangle$. Black dots are VNA measurements, colored dots show ring-down measurements for low (red) to high (orange) pulse powers.}
    \label{fig:Qint_aged}
\end{figure}

An additional measurement cycle of the niobium cavity after several months of being exposed to air showed a decreased $Q_\text{int}$ ranging from $0.26-0.38\times 10^9$, illustrated in \cref{fig:Qint_aged}. This indicates the presence of many more TLS, due to the grown oxide layer on the cavity wall. We are now not able to saturate the TLS with our available measurement equipment.  Ring-down measurements demonstrate a similar, but much less pronounced deviation from VNA measurements as presented in the main paper. 
A lower signal-to-noise ratio in the aged cavity time domain measurement cycle prevents precise characterizations with the TLS model. Still, a slight trend of an increased internal quality factor $Q_\mathrm{int}$ during the cavity ring-down, as predicted from the TLS model, can be confirmed. 

\subsection{Aluminium and copper cavities}
We first tested the coaxial quarter-wave resonator with three different purity grades of copper, 2N, 4N5, and 6N. The advantage of a normal conducting cavity would be the possibility to introduce magnetic fields, which is not possible with superconductors due to the Meissner effect. The copper cavities reach the lowest internal quality factors, as the magnetic participation ratio of the resonator design is not ideal for normal conducting metals. Both etching with cupric chloride and annealing in the presence of reduced oxide pressure improved the cavity’s internal quality factor to values in the range of $Q_\mathrm{int} \approx 10^4 $. The impact of copper purity on $Q_\mathrm{int}$ becomes negligible after one or both surface treatments.
Changing the material to high-purity aluminium (5N) leads to a quality factor of $Q_\text{int}=13\times10^6$, after a surface removal of more than $150 \, \mathrm{\mu m}$. A weak $Q_\mathrm{int}$ power dependence indicates the presence of saturable TLS on the cavity surface. Still, the main loss mechanism seems to be either quasiparticles with a finite surface resistance or a non saturable dielectric layer on top of the inner cavity walls.

\bibliography{references}

\end{document}